\documentclass[
aps,
prl,
reprint,
floatfix,
amsmath,
twocolumn,
superscriptaddress,
altaffillsymbol
]{revtex4-1}

\usepackage{dcolumn}   % needed for some tables
\usepackage{bm}        % for math
\usepackage{amssymb}   % for math
\usepackage{amsfonts}
\usepackage{verbatim}
\usepackage{mathrsfs}
\usepackage{graphicx}
\usepackage{amsmath}
\usepackage{amsfonts}
\usepackage{ulem}

\def\lsim{\mathrel{\rlap{\lower4pt\hbox{\hskip1pt$\sim$}}
    \raise1pt\hbox{$<$}}}                % less than or approx. symbol
\def\gsim{\mathrel{\rlap{\lower4pt\hbox{\hskip1pt$\sim$}}
    \raise1pt\hbox{$>$}}}                % greater than or approx. symbol

\begin{document}

\title{Reverse isolation and backaction of the SLUG microwave amplifier} %Title of paper

\author{T. Thorbeck}
\thanks{These authors contributed equally to this work. T. T. is currently at IBM T. J. Watson Research Center, Yorktown Heights, New York 10598, USA.}
%\altaffiliation[Present address: ]{IBM T. J. Watson Research Center, Yorktown Heights, New York 10598, USA}
\author{S. Zhu}
\thanks{These authors contributed equally to this work. T. T. is currently at IBM T. J. Watson Research Center, Yorktown Heights, New York 10598, USA.}
\author{E. Leonard Jr.}
\affiliation{Department of Physics, University of Wisconsin-Madison, Madison, Wisconsin 53706, USA}
\author{R. Barends}
\author{J. Kelly}
\affiliation{Google Inc., Santa Barbara, California 93117, USA}
\author{John M. Martinis}
\affiliation{Google Inc., Santa Barbara, California 93117, USA}
\affiliation{Department of Physics, University of California, Santa Barbara, California 93106, USA}
\author{R. McDermott}

%\author{T. Thorbeck}
%\thanks{These two authors contributed equally.}
%\altaffiliation[Present address: ]{IBM T. J. Watson Research Center, Yorktown Heights, New York 10598, USA}
%\affiliation{Department of Physics, University of Wisconsin-Madison, Madison, Wisconsin 53706, USA}
%%\affiliation{Department of Physics, University of Wisconsin-Madison, Madison, Wisconsin 53706, USA}
%\author{S. Zhu}
%\thanks{These two authors contributed equally.}
%\affiliation{Department of Physics, University of Wisconsin-Madison, Madison, Wisconsin 53706, USA}
%\author{E. Leonard Jr.}
%\affiliation{Department of Physics, University of Wisconsin-Madison, Madison, Wisconsin 53706, USA}
%\author{R. McDermott}
%\email[Electronic address: ]{rfmcdermott@wisc.edu}
%\affiliation{Department of Physics, University of Wisconsin-Madison, Madison, Wisconsin 53706, USA}

\email[Electronic address: ]{rfmcdermott@wisc.edu}
\affiliation{Department of Physics, University of Wisconsin-Madison, Madison, Wisconsin 53706, USA}
%\altaffiliation[Present address: ]{IBM T. J. Watson Research Center, Yorktown Heights, New York 10598, USA}

\date{\today}

\begin{abstract}

An ideal preamplifier for qubit measurement must not only provide high gain and near quantum-limited noise performance, but also isolate the delicate quantum circuit from noisy downstream measurement stages while producing negligible backaction. Here we use a Superconducting Low-inductance Undulatory Galvanometer (SLUG) microwave amplifier to read out a superconducting transmon qubit, and we characterize both reverse isolation and measurement backaction of the SLUG. For appropriate dc bias, the SLUG achieves reverse isolation that is better than that of a commercial cryogenic isolator. Moreover, SLUG backaction is dominated by thermal emission from dissipative elements in the device. When the SLUG is operated in pulsed mode, it is possible to characterize the transmon qubit using a measurement chain that is free from cryogenic isolators or circulators with no measurable degradation of qubit performance.

\end{abstract}


\maketitle %\maketitle must follow title, authors, abstract and \pacs

Fault-tolerant quantum computation in the surface code demands fast, high-fidelity measurement of multiqubit parity operators \cite{Fowler12}. Measurement involves monitoring a microwave signal that is transmitted across or reflected from a linear cavity that is dispersively coupled to the qubit. To achieve high fidelity, it is necessary that the noise contribution of the first-stage amplifier be close to the standard quantum limit. However, the demands of operating a large-scale superconducting processor require global optimization of the measurement chain, and amplifier added noise is but one consideration. The measurement system must isolate the qubit from the noise of downstream amplification stages at higher temperatures, while at the same time producing minimal classical backaction on the qubit, due either to stray microwave power at pump tones or to emission from dissipative elements. Finally, the measurement system must be designed with an eye to overall wiring simplicity and minimum system footprint.

In most superconducting qubit measurements, preamplification is provided by some form of Josephson parametric amplifier \cite{Castellanos07, Bergeal10, Hatridge11}. Such devices operate at or near the standard quantum limit; however, integration requires extensive use of nonreciprocal elements such as isolators or circulators, which rely on magnetic materials to break time reversal symmetry and achieve nonreciprocal transmission characteristics. Commercial ferrite-based isolators and circulators are bulky, magnetic, and expensive, so they are not a scalable technology. There have been prior attempts to engineer nonreciprocal gain in superconducting parametric amplifiers, notably using coupled Josephson parametric converters (JPCs) \cite{Abdo13, Abdo14}. However, the bandwidth and saturation power of such devices are quite limited, complicating efforts to perform multiplexed qubit readout. The Josephson traveling wave parametric amplifier (TWPA) \cite{Macklin15} and the kinetic inductance traveling-wave (KIT) amplifier \cite{Eom12} display directionality, but in the ideal case the reverse gain of these devices is 0 dB: signals coupled to the output port propagate unattenuated through to the input. There are ongoing efforts to engineer a Josephson circulator that can provide on-chip reverse isolation \cite{Kamal11, Kerchoff15, Sliwa15}; however, these have not yet demonstrated sufficient bandwidth to enable multiplexed qubit readout.

An alternative approach to qubit measurement involves low-noise preamplification using the Superconducting Low-inductance Undulatory Galvanometer (SLUG) \cite{Ribeill11, Hover12}, a variant of the dc Superconducting QUantum Interference Device (dc SQUID). Previous experiments have shown significant improvements in single-shot qubit measurement when the SLUG is employed as a first-stage amplifier \cite{Hover14}, and wireup and operation of the SLUG is particularly simple as the device requires only two dc current biases and no microwave pump tones. It has long been known that the SQUID provides intrinsic nonreciprocity \cite{Clarke79, Ranzani13b}, and recent theoretical studies explore SQUID directionality as a consequence of asymmetric frequency conversion involving upconversion to and downconversion from the Josephson frequency and harmonics, leading to efficient coupling of a differential-mode input signal to a common-mode output and suppression of the reverse process \cite{Kamal12, Ranzani15}.

In this Letter, we analyze the directionality of the SLUG microwave amplifier and show that reverse isolation can be extremely high, comparable to that achieved using commercial cryogenic isolators. For this reason, it is possible to integrate the SLUG into a qubit measurement chain that involves no cryogenic isolators or circulators. We perform readout of a transmon qubit using the SLUG and analyze classical backaction due to thermal emission from dissipative elements of the amplifier. For appropriate pulsed bias of the SLUG, it is possible to eliminate isolators and circulators from the qubit measurement chain with no measurable degradation of coherence.

In Figs. 1a-b we show a circuit diagram of the SLUG element along with a micrograph of the fabricated device. The Nb/AlO$_x$/Nb Josephson junctions were formed in 2~$\mu$m$^2$ vias etched in the upper SiO$_2$ wiring dielectric. The critical current per junction is $I_0$~=~20~$\mu$A, corresponding to a critical current density of 1~kA/cm$^2$, and the shunt resistance per junction is $R = 8 \, \Omega$. The mutual inductance between the input signal and the SLUG loop is $M$~=~$L$~=~6.7~pH, and the peak-to-peak voltage modulation of the device is around 130~$\mu$V. The input matching network is a single-pole lumped element $LC$ section with a designed characteristic impedance of 2~$\Omega$. The signal to be amplified is injected directly into the loop of the gain element as a current, and the flux through the loop induces a change in voltage at the SLUG output via the usual quantum interference action (Fig. 1c). The device is operated at a quasistatic bias point on the left or right shoulder of the $V$-$\Phi$ curve where the flux-to-voltage transfer coefficient $V_\Phi \equiv \partial V/\partial \Phi$ is large, so that a small change in input current yields a large change in output voltage; the sign of $V_\Phi$ can be either positive or negative. Reverse isolation is determined by the reverse transimpedance $Z_r \equiv \partial V_{in}/\partial I_{out}$. In Fig. 1d we plot Im[$Z_r$] calculated after the method described in \cite{Ribeill11}. The reverse transimpedance depends strongly on the sign of $V_\Phi$, as we discuss in detail below.%For power matching to a 50~$\Omega$ microwave source, the input signal is coupled to the SLUG element via an $LC$ or transmission-line impedance transformer.
\begin{figure}
\includegraphics[width=\columnwidth]{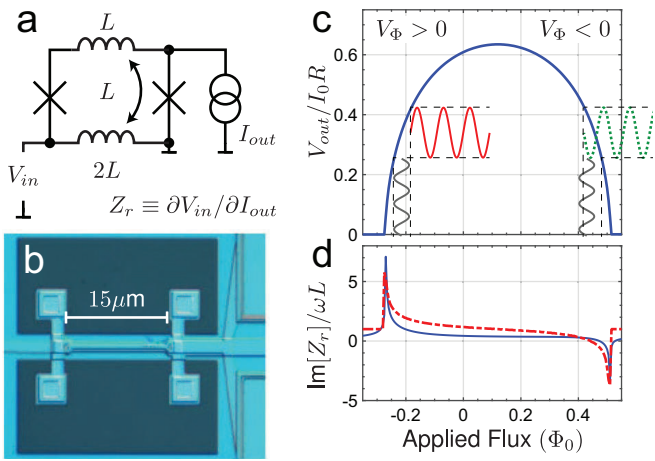}%

\caption{\label{Figure 1} (a) Circuit diagram of the SLUG. In typical operation, the input signal is coupled as a current to the node at the lower left, and the output is taken from the node at upper right. (b) Micrograph of SLUG gain element.  (c) SLUG flux-to-voltage ($V-\Phi$) transfer curve calculated after \cite{Ribeill11}. The sign of $V_\Phi$ changes depending on whether the device is biased on the left or right shoulder of the $V-\Phi$ curve. (d) Im$[Z_r]$ calculated after \cite{Ribeill11} (blue solid trace) and from the approximate expression $1+LV_\Phi/R$ (red dashed trace).}
\end{figure}

In Fig. 2a we show a block diagram of the measurement setup for characterization of SLUG forward and reverse gain. Drive tones are coupled to the SLUG \textit{via} directional couplers, and a cryogenic coaxial relay routes signals from the input and output ports of the device to a single cryogenic HEMT postamplifier. Reflection from the unbiased SLUG allows \textit{in situ} calibration of the device scattering parameters \cite{Ranzani13}. In Fig. 2b we show simulated forward and reverse scattering parameters for the device obtained using the approach outlined in \cite{Ribeill11}, and in Fig. 2c we plot the measured data. For bias at points where the transfer function $\left|V_\Phi\right|$ is large, we observe large forward gain; moreover, the forward gain is insensitive to the sign of $V_\Phi$. The reverse gain is much lower; in addition, the reverse gain is asymmetric, with lower reverse gain for bias on the right shoulder of the $V$-$\Phi$ curve where $V_\Phi<0$. In Fig. 2d we show 1D cuts in forward and reverse gain taken on the right side of the $V-\Phi$ curve. For appropriate flux bias, the device reverse gain is around -20~dB over a band of order 1~GHz. This level of reverse isolation is comparable to that achieved with commercial cryogenic isolators.
\begin{figure*}
\includegraphics[width=\textwidth]{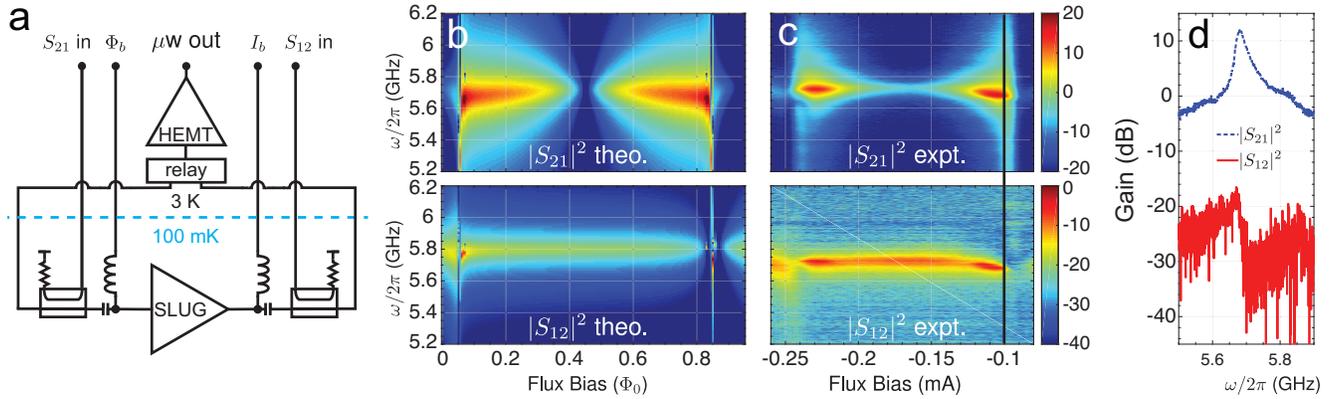}%
\caption{\label{Figure 2} SLUG forward and reverse gain. (a) Circuit for measurement of forward and reverse gain.   (b) Simulated forward and reverse scattering parameters as a function of flux bias and frequency.  (c) Measured forward and reverse scattering parameters. (d) 1D cuts from (c), showing forward and reverse gain versus frequency for large $\left|V_\Phi\right|$, $V_\Phi<0$.}
\end{figure*}

These results can be understood from a simple circuit model. For a device that is optimally power matched at the input and output ports, forward gain $S_{21}$ and reverse gain $S_{12}$ are given by the following expressions:
\begin{align}
\left|S_{21}\right|^2 &= \left|Z_f\right|^2/4R_iR_o \\
\nonumber
\left|S_{12}\right|^2 &= \left|Z_r\right|^2/4R_iR_o,
\end{align}
where $Z_f$ and $Z_r$ are the forward and reverse transimpedance of the SLUG, respectively, and $R_i = \rho_i (\omega L)^2/R$ and $R_o = \rho_oR$ are the real parts of the SLUG input and output impedance, respectively; here, $\rho_{i,o}$ are dimensionless, bias-dependent constants \cite{Ribeill11}.

The forward transimpedance is predominantly real, and is given by
\begin{align}
Z_f &= LV_\Phi.
\end{align}
In contrast, there are two contributions to reverse transimpedance. A current applied to the output node of the SLUG induces a voltage at the input node via Faraday induction, contributing a term to the reverse transimpedance of order $j\omega L$. At the same time, this current produces a voltage at the output node \textit{via} quantum interference; the SLUG inductance and junction dynamic resistance yield a voltage division, resulting in a second contribution to the reverse transimpedance of order $L V_\Phi \left(j \omega L/R\right)$. Importantly, this quantum interference contribution changes sign depending on which side of the $V_\Phi$ curve the device is biased. The two contributions to reverse transimpedance add coherently, so that
\begin{align}
\rm{Im}\left[\it{Z_r}\right] &= \chi_r \omega L \left( 1 + \frac{L}{R}\,V_\Phi \right),
\end{align}
where $\chi_r$ is a bias-dependent constant of order unity. For an optimized SLUG, we have $V_\Phi \approx R/L$ \cite{Ribeill11}; as a result, the Faraday and quantum interference contributions to reverse transimpedance are of the same order, and for $V_\Phi<0$ there exist bias points that provide excellent cancellation of these terms over a broad range of frequency.

Combining Eqs. 1-3, we can reexpress the forward and reverse gains as follows:
\begin{align}
\left|S_{21}\right|^2 &= \frac{1}{4\rho_i \rho_o}\left(\frac{V_\Phi}{\omega}\right)^2 \\
\nonumber
\left|S_{12}\right|^2 &= \frac{\chi_r^2}{4\rho_i \rho_o}\left( 1 + \frac{L}{R}\,V_\Phi \right)^2.
\end{align}
As a result, we find a SLUG directionality $\mathcal{D} \equiv \left|S_{21}\right|^2/\left|S_{12}\right|^2$ that is given by
\begin{align}
\mathcal{D} = \chi_r^{-2} \left(\frac{V_\Phi}{\omega}\right)^2 \left( 1 + \frac{L}{R}V_\Phi\right)^{-2}.
%\left(\frac{V_\Phi}{\omega}\right)^2 \bigg/ \chi_r^2 \left( 1 + \frac{L}{R}V_\Phi\right)^2.
\end{align}
For $V_\Phi \sim R/L > 0$, the last term in this expression is of order unity; as a result, we find $\mathcal{D} \sim \left(V_\Phi/\omega\right)^2$. For typical devices, we have $V_\Phi \approx$ 1 mV/$\Phi_0$, $V_\Phi/2\pi \approx$~80 GHz, so that for a device operating at 6~GHz, directionality is of order 20 dB. In contrast, for $\left|V_\Phi\right| \sim R/L, V_\Phi < 0$, the Faraday and quantum interference terms in the last term of Eq. 5 cancel, and directionality can be significantly better.

We have shown that the SLUG provides a level of reverse isolation comparable to or better than that of a commercial cryogenic isolator. However, the SLUG is operated in the finite voltage state and incorporates dissipative normal metal elements. Noise emission from the SLUG at the qubit frequency can cause spurious excitation or relaxation of the qubit \cite{Schoelkopf03, Martinis03}, although the qubit readout cavity will filter this noise. Noise emitted at the cavity frequency will populate the readout cavity with photons, and these photons will both shift the qubit frequency due to the ac Stark effect and induce photon shot-noise dephasing \cite{Schuster05, Sears12}. In the dispersive limit, the qubit and the cavity interact via the Hamiltonian $H_{int} = \hbar \chi \hat{n} \hat{\sigma_z}$, where $\chi$ is the strength of the qubit-cavity dispersive interaction, $\hat{\sigma_z}$ is the Pauli-$z$ operator, and $\hat{n}$ is the photon number operator for the resonator.  Finally, photons in the cavity can combine with noise at the qubit-cavity detuning frequency to cause spurious excitation or relaxation at the qubit frequency via dressed dephasing \cite{Boissonneault09, Slichter12}.

We have performed Ramsey interferometry experiments to quantify the level of classical backaction of the SLUG on an Xmon qubit circuit. The qubit is tunable over a frequency range from around 5 to 6 GHz, and the readout cavity mode is at 6.605~GHz. The experimental pulse sequence is shown in Fig. 3a. The SLUG idles in the supercurrent state and a fast flux pulse biases the SLUG into the active region prior to the initial $\pi/2$ pulse of a Ramsey fringe experiment; the SLUG ``head start'' time is varied from 0 to 1~$\mu$s. In Fig. 3b we show measured Ramsey fringes as a function of the SLUG head start time. Spurious photon population in the resonator affects qubit free evolution in two ways:  the precession frequency increases due to the ac Stark effect, and the dephasing rate increases due to shot noise fluctuations of the resonator photon occupation.  Both of these effects can be seen in the data of Fig. 3b. The initial rise in qubit precession frequency is on a timescale consistent with the cavity decay time, separately measured to be $1 / \kappa$~=~350~ns. The shift in qubit frequency saturates at 2.2~MHz for long SLUG head start times.  By comparing this shift in frequency to the separately measured qubit state-dependent cavity shift $2 \chi / 2 \pi$~=~1.5 MHz,  we calculate that this qubit frequency shift corresponds to a mean photon occupation of 1.5 in steady state (Fig.~3c).  %We limit the Ramsey precession time to 100 ns, less than the cavity decay time, to minimize the probability of a change in photon occupation during the experiment.  However, for short SLUG headstart times, the qubit frequency is changing during the course of the measurement, which could result in systematic errors in the qubit shift frequency and photon occupation.

\begin{figure}
\includegraphics[width=\columnwidth]{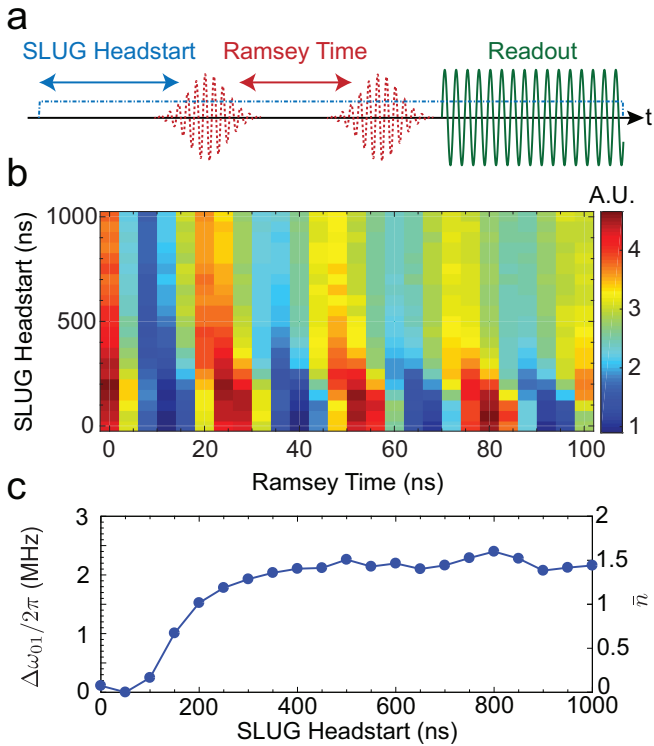}%
\caption{\label{Figure 3}  (a) Pulse sequence for characterizing SLUG backaction. A flux pulse is used to bias the SLUG into the active state prior to the initial $\pi/2$ pulse of a Ramsey sequence; the SLUG head start time is varied from 0 to 1~$\mu$s. (b) Qubit Ramsey fringes \textit{versus} free evolution time and SLUG head start time. (c) Mean photon occupation of the qubit readout resonator extracted from the data in (b).}
\end{figure}

We now consider the origin of the broadband emission from the SLUG. When the SLUG is biased in the active state, the circuit undergoes Josephson oscillations at a frequency that is well outside the signal band, typically around 40 or 50~GHz. Because these oscillations are far above the passband of the cabling and other microwave components in our measurement circuit and far from any qubit and cavity resonances, the qubit chip is likely protected from spurious emission at the SLUG Josephson frequency and harmonics. Another potential source of noise is thermal emission from the resistive shunts of the SLUG junctions. Static power dissipation in the SLUG shunts is around 1~nW. As the Pd thin-film shunt resistors occupy a relatively small volume, cooling of the electrons in the normal metal shunts is inefficient, and the electronic system equilibrates at a temperature that is far from the bath temperature of the cryostat. We can relate the electron temperature $T_e$ to the dissipated power $P$ as follows:
\begin{align}
\frac{P}{\Sigma V} = T_e^5- T_p^5,
\end{align}
where  $V$ is the volume of the shunt resistor, $\Sigma$ is the electron-phonon coupling constant, and $T_p$ is the phonon temperature \cite{Wellstood94}.  For $P$~=~1~nW, $V$~=~$5\times10^{-19}$~m$^3$, and $\Sigma = 1.2\times10^9$~W~m$^{-3}$~K$^{-5}$ \cite{Falferi08}, we find $T_e \approx$~1.1~K.  Since the SLUG input is well-matched to its 50~$\Omega$ environment, the output port of the qubit readout resonator sees an effective temperature of $T_e$ in the absence of any cryogenic isolators between the qubit and SLUG chips. For the multiplexed Xmon chip used in these measurements, coupling at the input and output ports is symmetric, yielding an effective temperature of the qubit readout resonator around 0.6 K, corresponding to a mean photon occupation of 1.5 for the 6.605 GHz mode. The agreement of this estimate with the measured average number of photons in the resonator suggests that thermal emission from the shunt resistors is the dominant source of classical backaction from the SLUG to the qubit circuit.

\begin{figure}
\includegraphics[width=\columnwidth]{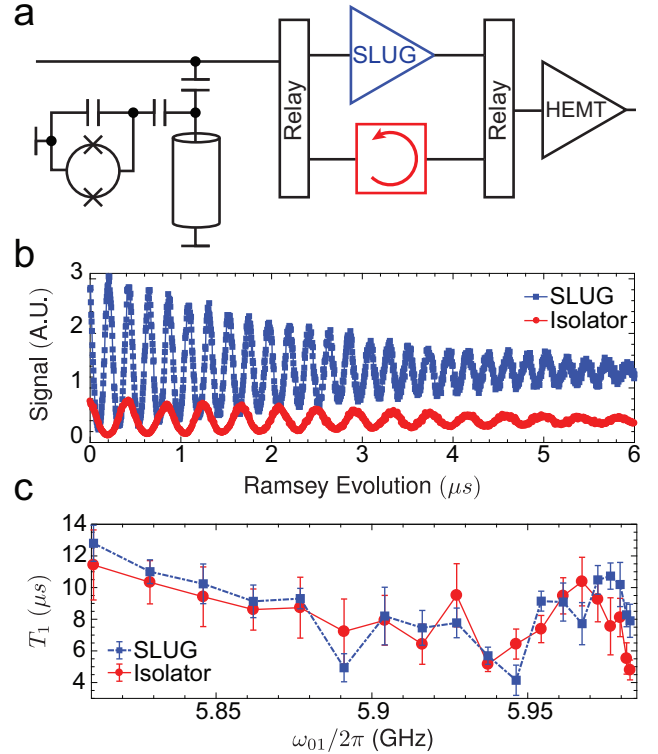}%
\caption{\label{Figure 4} (a) Circuit used to compare performance of pulsed SLUG to a single-stage cryogenic isolator. (b) Qubit Ramsey fringes measured in the two configurations. (c) Qubit $T_1$ times measured in the two configurations over a range of qubit frequency.}
\end{figure}

To circumvent this backaction, it is possible to operate the SLUG in pulsed mode so that it idles in the supercurrent state during qubit operation. The SLUG output then presents a superconducting short to ground, reflecting all broadband noise traveling upstream from the HEMT toward the quantum circuit. In a separate experiment, we used the SLUG to characterize an Xmon qubit with a measurement chain that is completely free from cryogenic isolators or circulators; the measurement circuit is shown in Fig. 4a. We employ two cryogenic coaxial relays that allow us to switch between a SLUG preamplifier and a single-stage cryogenic isolator between the qubit chip and the cryogenic HEMT amplifier; when the SLUG is switched in, there are no isolators or circulators in the measurement chain. In Fig. 4b we show qubit Ramsey fringes obtained with a conventional measurement chain (isolator + HEMT) and with the SLUG preamplifier (no isolator). Qubit coherent oscillations are monitored with a typical Ramsey sequence; upon the second $\pi/2$ pulse, the SLUG is flux biased to a point where $\left|V_\Phi\right|$ and forward gain are large. In each case, the heterodyne amplitudes were obtained by averaging 2000 traces with an integration time of 2 $\mu$s. SLUG preamplification yields a significant improvement in SNR; at the same time, we observe no degradation of qubit coherence. Finally, we have characterized qubit energy relaxation over a range of bias points for the two measurement configurations; results are shown in Fig. 4c. Within the error of the measurement, we observe no degradation of qubit energy relaxation when the cryogenic isolator is replaced in favor of the SLUG amplifier.

To conclude, we have characterized the reverse isolation and backaction of a nonreciprocal near quantum-limited linear amplifier based on the SLUG, a variant of the dc SQUID. As with any SQUID, the device is nonreciprocal; however, in the case of the SLUG the destructive interference of two contributions to reverse transimpedance yields exceptional directionality over a wide range of frequency. We have characterized classical backaction of the SLUG and shown that it is compatible with hot electron effects; integration of large-volume cooling structures is expected to lower the electron temperature. By operating the device in pulsed mode, it is possible to circumvent backaction and to realize a measurement chain that is entirely free from cryogenic isolators or circulators. As a result, the SLUG offers a path to scalable high-fidelity readout in large-scale quantum processors where wiring footprint and complexity must be considered in the overall system optimization.

\begin{acknowledgments}
% Put your acknowledgments here.
This work was supported by the U.S. Government under Grant W911NF-14-1-0080. Portions of this work were performed in the Wisconsin Center for Applied Microelectronics, a research core facility managed by the College of Engineering and supported by the University of Wisconsin - Madison.
\end{acknowledgments}


\end{document}